%% file: paper.tex
\documentclass[10pt,twocolumn]{article}
\usepackage{times}
\usepackage{fixltx2e}
\usepackage{graphicx}
\usepackage{longtable}
\usepackage{float}
\usepackage{wrapfig}
\usepackage{rotating}
\usepackage[normalem]{ulem}
\usepackage{amsmath}
\usepackage{textcomp}
\usepackage{marvosym}
\usepackage{wasysym}
\usepackage{amssymb}
\usepackage{hyperref}
\usepackage{color}
\usepackage{alltt}
\usepackage{xspace}
\usepackage{verbatim}
\usepackage{csquotes}

\input{macros}

\begin{document}

\title{A Graphical Interactive Debugger for Distributed Systems}
\author{Doug Woos$^1$ and Zachary Tatlock$^1$ and Michael D. Ernst$^1$ and Thomas E. Anderson$^1$ \\
\small {\em  $^1$University of Washington} \\ [2mm]
\small Submission Type: Research
}
\date{}
\maketitle

\input{abstract}
\input{intro}
\input{overview}
\input{frontend}
\input{backend}
\input{implementation}
\input{eval}

\input{discussion}
\input{related}
\input{conclusion}

\bibliographystyle{abbrv}
\bibliography{bibliography}

\end{document}

%% file: macros.tex
\newcommand{\todo}[1]{\relax}

\newcommand{\para}[1]{\textbf{#1.}}

\newcommand{\ie}{i.e.\xspace}

\newcommand{\elidedtext}[1]{}

\def\|#1|{\mathid{#1}}
\newcommand{\mathid}[1]{\ensuremath{\mathit{#1}}}
\def\<#1>{\codeid{#1}}
\protected\def\codeid#1{\ifmmode{\mbox{\smaller\ttfamily{#1}}}\else{\smaller\ttfamily #1}\fi}


%% file: abstract.tex
\begin{abstract}
Designing and debugging distributed systems is notoriously difficult.
The correctness of a distributed system is largely determined by its handling of
failure scenarios. The sequence of events leading to a bug can be long and
complex, and it is likely to include message reorderings and failures. On
single-node systems, interactive debuggers enable stepping
through an execution of the program, but they lack the ability
to easily simulate failure scenarios and control the order in which messages are
delivered.

Oddity is a graphical, interactive debugger for distributed systems. It brings
the power of traditional \emph{step-through debugging}---fine-grained control and
observation of a program as it executes---to distributed systems.
It also enables \emph{exploratory testing}, in which an engineer
examines and perturbs the behavior of a system in order to
better understand it, perhaps without a specific bug in mind.
A programmer can directly control message and failure
interleaving.
Oddity supports \emph{time travel}, allowing a developer to explore
multiple branching executions of a system within a single debugging
session.
Above all, Oddity encourages distributed systems thinking: rather than
assuming the normal case and attaching failure handling as an afterthought,
distributed systems should be developed around the certainty of message
loss and node failure.

Graduate and undergraduate students used Oddity in two distributed systems
classes.  Usage
tracking and qualitative surveys showed that students found Oddity
useful for both debugging and exploratory testing.
\end{abstract}
%


%% file: intro.tex
\section{Introduction}
Developing correct distributed systems is difficult. Such systems are
inherently nondeterministic.
Messages can be dropped or arbitrarily delayed, and nodes can fail and
restart.
In the ``normal'' case where messages are delivered in order and
nodes remain up and responsive, understanding the behavior of the code,
as well as testing and debugging, are all relatively simple.
However, bugs are more likely to hide in the unusual failure cases.
For example, a version of the widely-used Raft consensus algorithm
\cite{DBLP:conf/usenix/OngaroO14} was discovered to have a bug in the
code to handle changes in the participants to the protocol,
depending on the interleaving of reconfiguration requests and
a leader failover.

For single-node systems, engineers have step-through debuggers. A debugger helps
an engineer reproduce and understand bugs by observing how their
system's state evolves in both normal and buggy executions. However, traditional
interactive debuggers are of limited utility in debugging distributed systems:
they do not allow programmers to easily control which messages will be delivered
and in what order. Since the behavior of a distributed system is determined by
the order in which events happen, engineers cannot debug their systems without
this control. Even the simple sanity checks that engineers can do with a
traditional debugger in order to ensure that they understand how their system
operates (e.g., for a given input, how many times is the inner loop executed?)
are out of reach in a distributed system.

To address this, we present Oddity, a graphical, interactive debugger for
distributed systems.\footnote{Oddity is open source and available at
  \url{http://oddity.uwplse.org}.} It enables engineers to explore and control the
execution of their system, including both normal operation and edge
cases---message drops, node failures, and delays. Oddity displays the
messages and timeouts that are waiting to be delivered and allows the engineer
to specify their order.  Oddity supports
time-travel, allowing the engineer to navigate a branching history of possible
executions.  This enables users to backtrack and make different choices about
the order in which messages and timeouts are delivered, allowing the exploration
of many different cases---for instance, all of the possible orderings of a few
messages---in a single debugging session. By enabling programmers to easily
explore both normal cases and edge cases---indeed, Oddity makes no
distinction between these cases---Oddity encourages distributed systems
thinking.  Rather than assuming the normal case and attaching failure handling
as an afterthought, systems should be developed around the possibility of failure
and then optimized for performance.

Oddity supports a general execution model: event handlers run in response to
received messages or timeouts. A handler can modify local state, set timeouts,
and send messages to other nodes.  Handlers can be written in any programming
language, needing only to support a simple shim API for interaction with the
debugger (sending and receiving messages, setting timeouts, and updating the
node's state).

Oddity differs from previous work in several ways. Unlike previous distributed
systems visualization tools, it can be used to visualize and control the network
behavior of a real system, developed in any programming language. Other systems
only visualize the operation of a model~\cite{runway} or logs of a particular
execution~\cite{shiviz}. Similarly, previous debugging systems for distributed
systems~\cite{Bates:1983:AHD:800007.808017,Eick:1996:IVM:525394.837855,Kranzlmuller:1996:EGV:238020.238054,380478,Kunz97poet:target-system-independent,Zernick:1992:UVT:624593.625178,d3,shiviz} focused on ex post facto debugging and diagnosis, while Oddity is geared
toward interactive exploration of executions.
Oddity is extensible and supports multiple representations for viewing or
interacting with a distributed system.
Oddity supports two visual representations of a system execution.
One, shown in Section~\ref{sec:overview}, emphasizes the current state of the
nodes and the network, including in-flight messages and timeouts, and also
enables navigation through an execution.
To demonstrate Oddity's extensibility, we added a traditional
(non-interactive) space-time diagram representation in under 150 lines of code.

This research makes the following high-level contributions:

\begin{itemize}
\item Interactive debugging of distributed systems. Oddity is the first system
  that allows users to interactively control the order of messages and timeouts
  that are delivered to each node in a distributed system, enabling both
  debugging and exploratory testing.  Oddity is designed to encourage and enable
  users to reason about the correctness of their systems by exploring edge cases
  as well as normal cases. Developers can use Oddity to visualize execution
  traces mined from logs or obtained from a model-checker as a counterexample
  to a desired invariant.

\item A conceptual model for distributed systems development. In Oddity, all
  messages and timeouts for a given node are grouped together in ``inboxes,''
  indicating that any event in any inbox can occur at its node at any time, and that systems
  cannot assume a ``normal'' ordering. Oddity models event history as a tree of
  possible executions, allowing a programmer to explore the consequences of
  various event orderings by navigating multiple executions of their system.

\item A novel graphical interface. Oddity includes a new graphical
  representation of the partial execution of a distributed system, designed to
  encourage users to think carefully about the correctness of their
  systems. This interface allows users to inspect a single state of the system
  in detail, while also enabling navigation through an execution of the system.

\item A study of student usage of Oddity. Students used Oddity in lab
  assignments for two distributed systems classes. We studied
  students' experiences with Oddity with opt-in usage tracking and an optional
  survey. In addition to providing evidence that Oddity is useful for
  distributed systems development, our classroom experiment provides the first
  insights into student behavior with an interactive debugger for distributed
  systems.
\end{itemize}

The rest of the paper is organized as follows. Section~\ref{sec:overview}
presents Oddity from a user's perspective via a running example: diagnosing a
bug in the Raft consensus protocol. Section~\ref{sec:frontend} discusses
Oddity's graphical interface in more detail. Section~\ref{sec:backend} discusses
Oddity's architecture. Section~\ref{sec:implementation} discusses our prototype
implementation of Oddity. Section~\ref{sec:eval} discusses our deployment of
Oddity to two university distributed systems classes; students were able to use
Oddity for both debugging and exploratory testing. Section~\ref{sec:related}
discusses related work, and Section~\ref{sec:conclusion} concludes and presents some
potential avenues for future work.

%% file: overview.tex
\section{Example usage}\label{sec:overview}

We introduce Oddity's core ideas and interface via a running example: an
implementation of Raft~\cite{DBLP:conf/usenix/OngaroO14}.  Raft is a consensus
protocol, a key component in the construction of strongly-consistent distributed
services. A consensus protocol enables a cluster of nodes to agree on a sequence
of values to apply to a state machine, despite node failures and arbitrary
message delays.  To support changes in the nodes participating in the state
machine consensus, Raft includes a reconfiguration protocol in which both the
new and old sets of nodes must agree on any new configuration. The
reconfiguration protocol could be triggered manually by a system administrator
or automatically by a cluster management system.  In part because it was
well-documented and included source code, Raft has become widely deployed in
industry.

Ongaro's dissertation~\cite{ongaro:phd}
includes a simplified reconfiguration protocol designed for single node changes.
Several years after publication, researchers discovered a bug in this simplified
protocol: in a cluster with an even number of members, if two competing
reconfiguration requests occur with a leader election in between, the cluster
can lose data. A simple fix, proposed when Ongaro publicly announced the bug, is
to require that new leaders commit an entry to the log in the old configuration
before committing a new configuration. Several months passed between
the bug being identified and the fix being announced.

For explanatory purposes, we imagine a Raft maintainer has been informed
of the existence of the buggy execution; using Oddity, they are trying
to determine why it happens and how it can be fixed.

\begin{figure}
\centering
\includegraphics[width=0.95\linewidth]{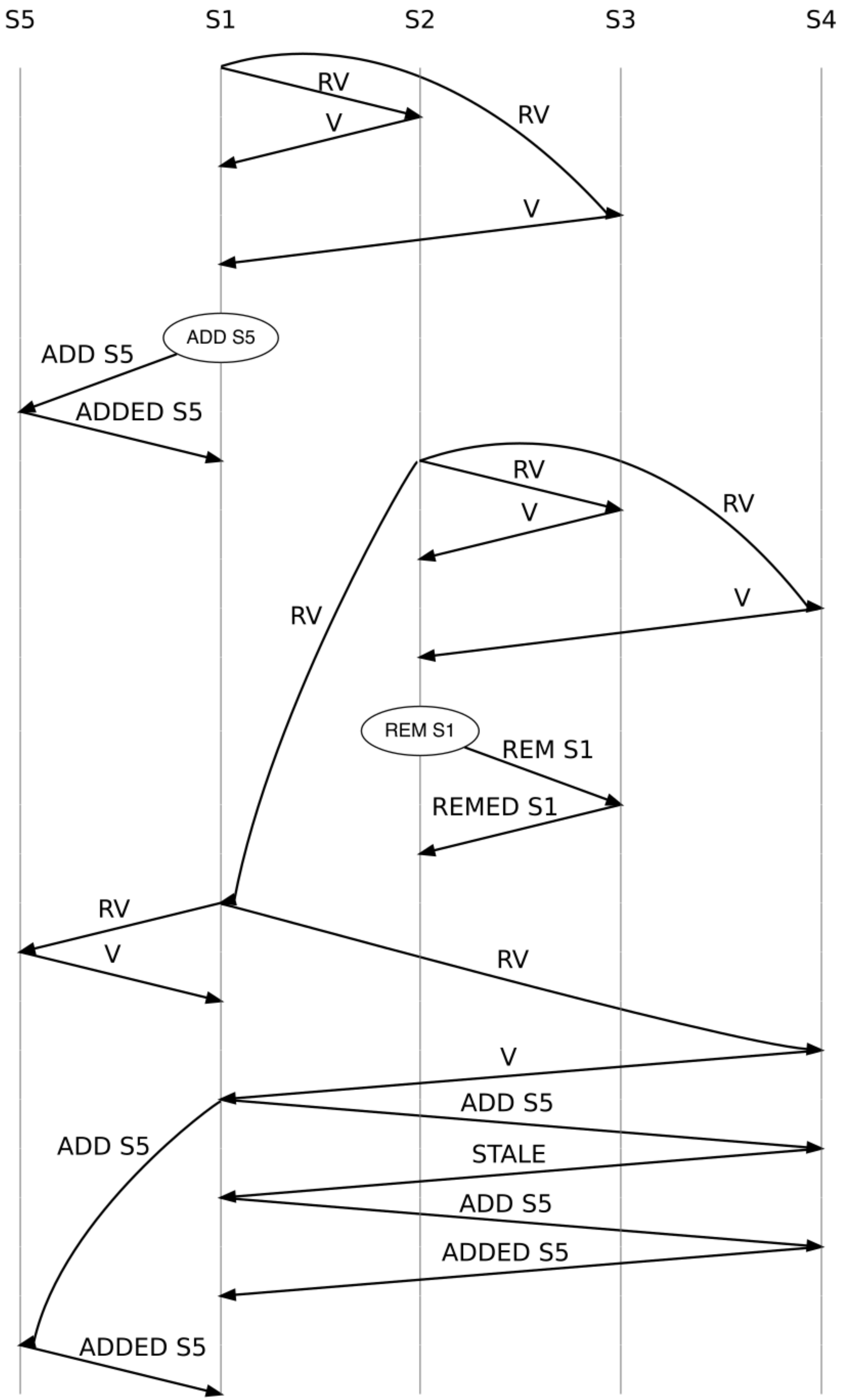}
\caption{ A space-time diagram, generated with Oddity and lightly edited for
  clarity, illustrating a Raft execution leading to the reconfiguration
  bug. Each vertical line represents in a node in the system, and arrows between
  them represent messages. The circles represent messages received from clients
  of the system (elided for presentation).}
\label{fig:raft-bug}
\end{figure}

Figure~\ref{fig:raft-bug} shows
an execution leading to the Raft bug.
First, a leader ($S_1$) is elected in a 4-node
cluster by a majority including itself, $S_2$, and $S_3$. Then, $S_1$ starts to replicate a new configuration that adds a
fifth node, $S_5$. In the single-server reconfiguration protocol, each server uses
whichever configuration is latest in its log (regardless of whether it is
committed).  The leader sends this new configuration to $S_5$
(shown on the left of Figure~\ref{fig:raft-bug}) as well as
the rest of the cluster (assumed to be delayed or dropped in Figure~\ref{fig:raft-bug}). After this configuration is replicated to $S_5$, $S_2$
starts an election and is elected with votes from $S_3$ and $S_4$.
This might occur, for example, if the reconfiguration messages from $S_1$ are
delayed to those nodes, e.g., due to a temporary network
outage. (Consensus should work even when nodes
incorrectly judge that other nodes have failed.)  Now that
$S_2$ is leader, it starts to replicate a new configuration that removes $S_1$
from the cluster, leaving the three nodes $S_2$, $S_3$, and $S_4$ (since the configuration with
$S_5$ was never replicated to $S_2$). It successfully replicates this
configuration to $S_3$, at which point it can commit the configuration since it is on a
majority of nodes in the new cluster of three nodes. Now $S_1$ starts another election, and
becomes leader with votes from $S_4$ and $S_5$. It can now finish replicating
its configuration adding $S_5$ to the whole cluster, which overwrites $S_2$'s
committed configuration. This is a violation of a crucial Raft safety property:
once an entry is committed, it should never be overwritten.

Without Oddity, the Raft engineer has several options to reproduce and diagnose
this failure. They could examine the code and try to imagine an execution that
would trigger the bug, but this is both time-consuming and error-prone. They
could design an automated test to find the issue, but testing distributed systems
is notoriously difficult~\cite{Maddox:2015:TDS:2800695.2800697}. Since the issue
depends on a failover, the test environment would need to simulate a temporary
network partition. The test environment would also need to ensure that messages
are delivered in a specific order with respect to other messages and the network
outage. The engineer would also need to write an oracle that determines whether
the bug has in fact been triggered (\ie, whether data are lost). Finally,
the engineer could run their code in a traditional debugger and attempt to
trigger the issue. Doing so, however, would still require simulation of failures
and control over the order in which messages are delivered.

The rest of this section shows how Oddity makes the Raft engineer's task
easier.\footnote{A screen-cast version can be found at
  \url{http://oddity.uwplse.org}.}  This illustrates Oddity's functionality via
one important use case: reproducing and diagnosing a bug in a distributed
system. Oddity can also be used for open exploration of a distributed system's
behavior, or for visualizing a counterexample produced from a model checker.

\subsection{Initialization}

Oddity assumes that the system being debugged is implemented as a set of event
handlers: deterministic functions that can read and write the node's state,
send messages, and set timeouts through the Oddity shim API (detailed in
Section~\ref{sec:backend}).  Oddity communicates with the shim, which calls
event handlers.

In order to use Oddity, the engineer makes a few changes to the system.
The engineer first connects the system's handlers to
Oddity by routing communication through the Oddity shim instead of the standard
networking library (this can typically be achieved by a small macro or command
line flag).
Next, the engineer creates a node to represent the system's client.
This node can set timeouts that
cause communication with the rest of the system. In response to a
timeout, the Raft client should send the reconfiguration commands from the
counterexample in Figure~\ref{fig:raft-bug}. The client can be developed using
any language for which an implementation of the Oddity shim is available.

Having linked the system and the client with the shim, the engineer can run
the system under Oddity.

\subsection{Finding a buggy execution}

\begin{figure}
\centering
\fbox{\includegraphics[width=.9\linewidth]{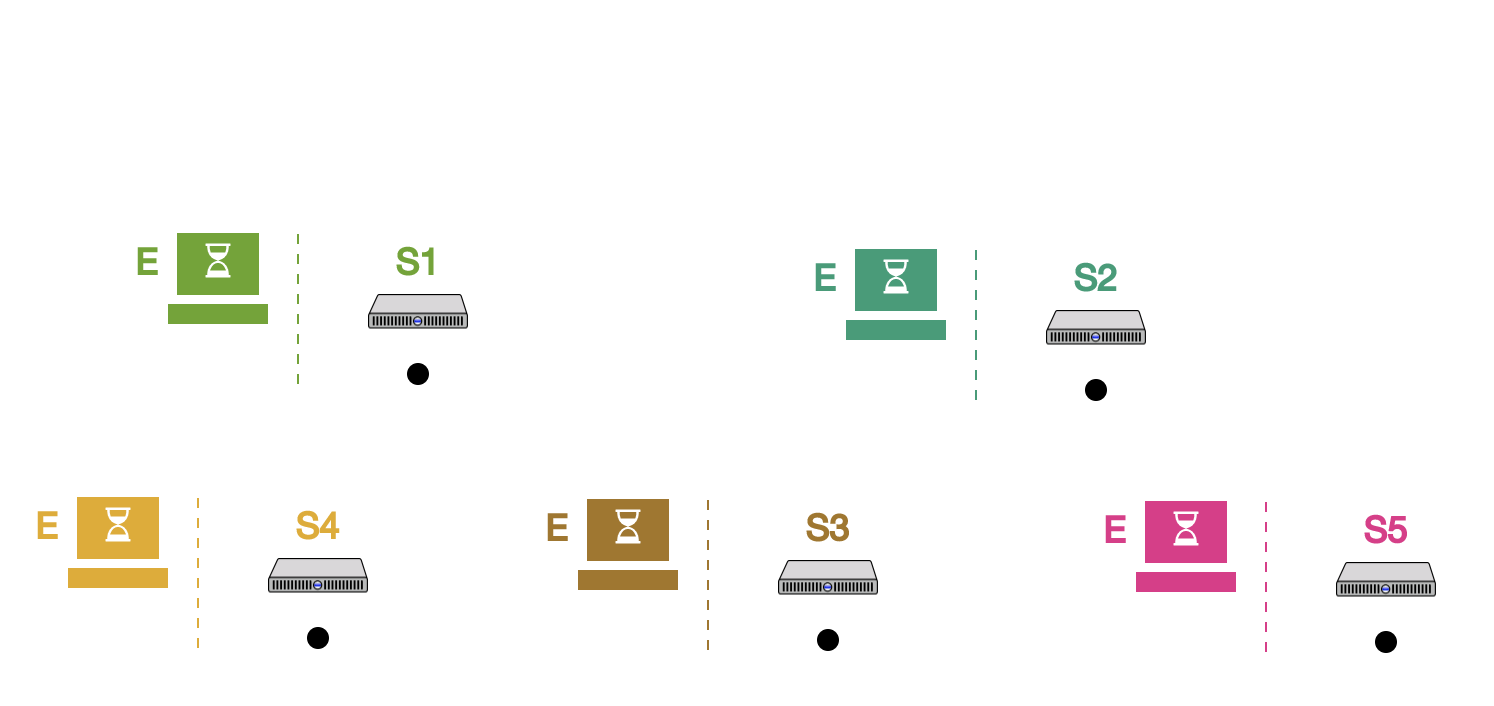}}
\caption{\label{fig:raft-1}The initial state of the Raft system in Oddity. Each
  node has a timeout in its inbox.}
\end{figure}

When the engineer first starts Oddity on their system, they will see a screen
similar to Figure~\ref{fig:raft-1}. Each node has an ``inbox'' next to it, which
contains both messages sent by other nodes and also timeouts the node has set
itself.  At the beginning of time, no messages have been sent, so each node's
inbox contains only the timeouts waiting at that node (including $S_5$, which
has not yet been added to the cluster).  Timeouts are used to cause events to
fire without messages being delivered. For instance, the election timeouts in
Figure~\ref{fig:raft-1} are fired when a node has not received a message from a
leader for sufficient time, and cause the node to start an election.

\begin{figure}
\centering
\fbox{\includegraphics[width=.9\linewidth]{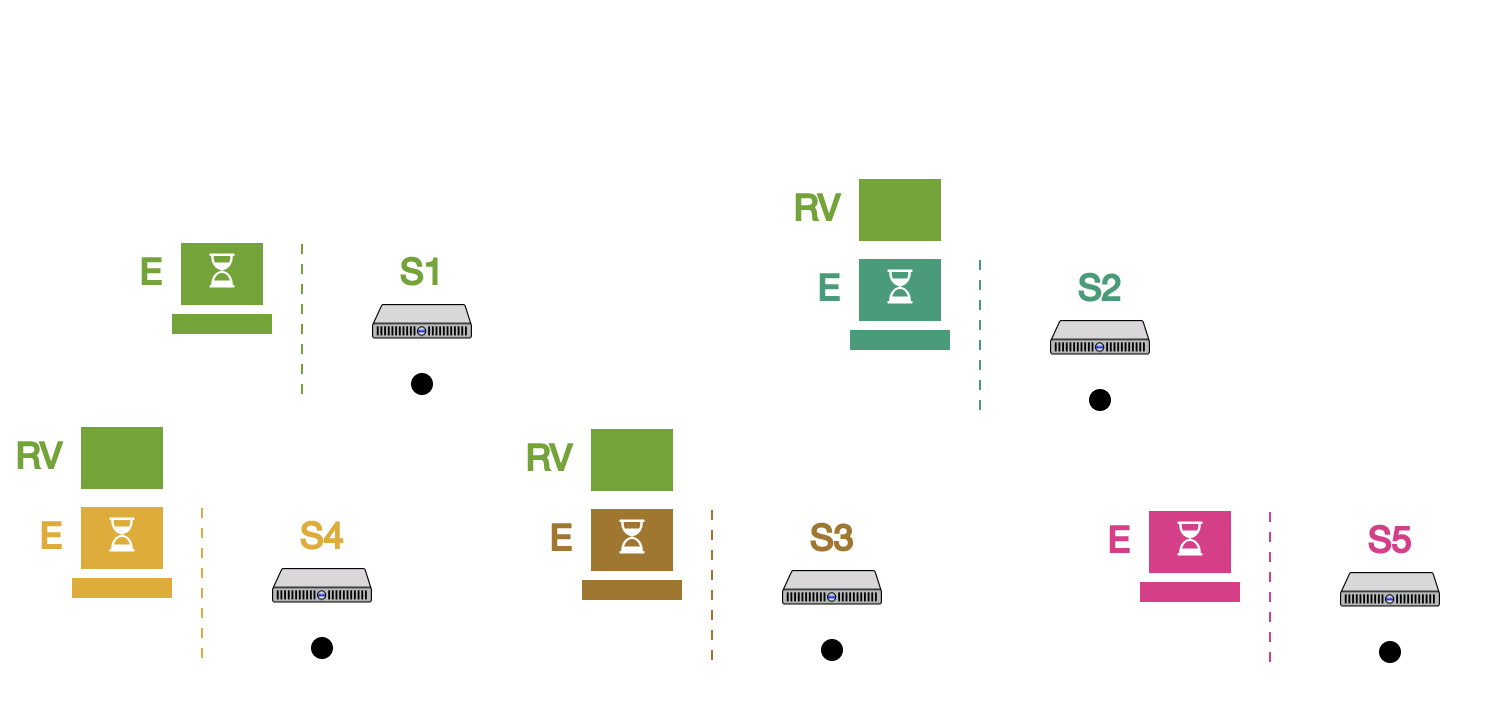}}
\caption{\label{fig:raft-2}The state of the Raft system in Oddity after $S_1$
  starts an election. $S_2$, $S_3$, and $S_4$ have \texttt{RV} messages in their
inboxes. The messages have the same color as their sender ($S_1$).}
\end{figure}

The engineer will first need to get $S_1$ elected leader. They can click on the
\texttt{E} (election) timeout in $S_1$'s inbox to deliver it, causing $S_1$ to
send \texttt{RV} (Request Vote) messages to the other nodes in the initial
configuration (excluding $S_5$, which has not yet been added). The resulting
state of the system, with a \texttt{RV} message in each node's inbox,
is shown in Figure~\ref{fig:raft-2}. These messages are now waiting to be
delivered.

\begin{figure}
\centering
\fbox{\includegraphics[width=.9\linewidth]{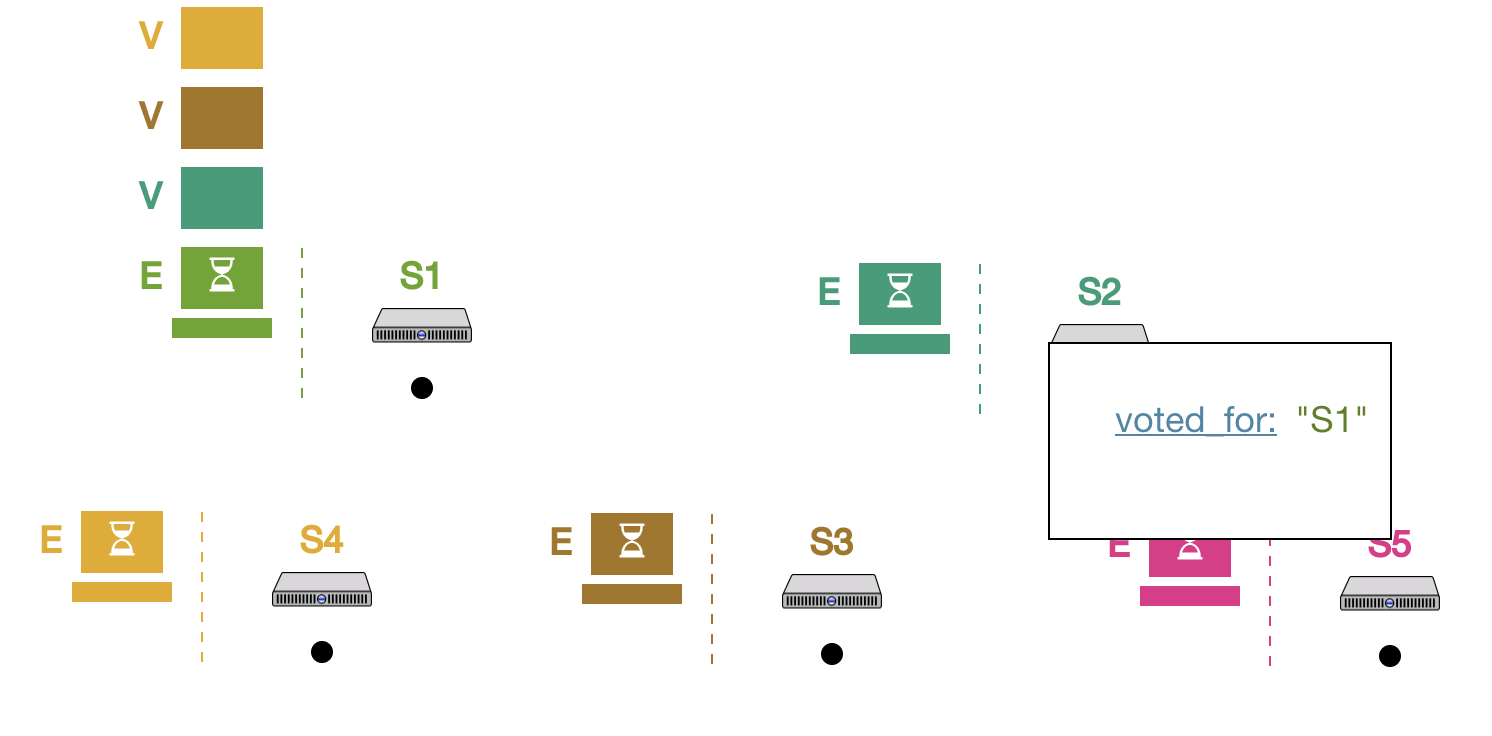}}
\caption{\label{fig:raft-3}The state of the Raft system in Oddity after $S_2$,
  $S_3$, and $S_4$ respond to $S_1$'s vote request. The votes from those nodes
  are in $S_1$'s inbox, with colors corresponding to the sending node. The
  engineer has clicked on $S_2$ to expand its state, showing that it has
  recorded a vote for $S_1$ (the Raft protocol requires that nodes track which
  node they voted for in the current term).}
\end{figure}

The engineer can then click on each \texttt{RV} message to deliver them, causing
them to respond to $S1$ with their \texttt{V} (Vote) messages. In
Figure~\ref{fig:raft-3}, these messages have been sent and $S_2$'s state is
expanded, showing that it voted for $S_1$. Since Raft requires a quorum to elect
a leader, once two of these votes are delivered $S_1$ considers itself elected.

\begin{figure}
\centering
\fbox{\includegraphics[width=.9\linewidth]{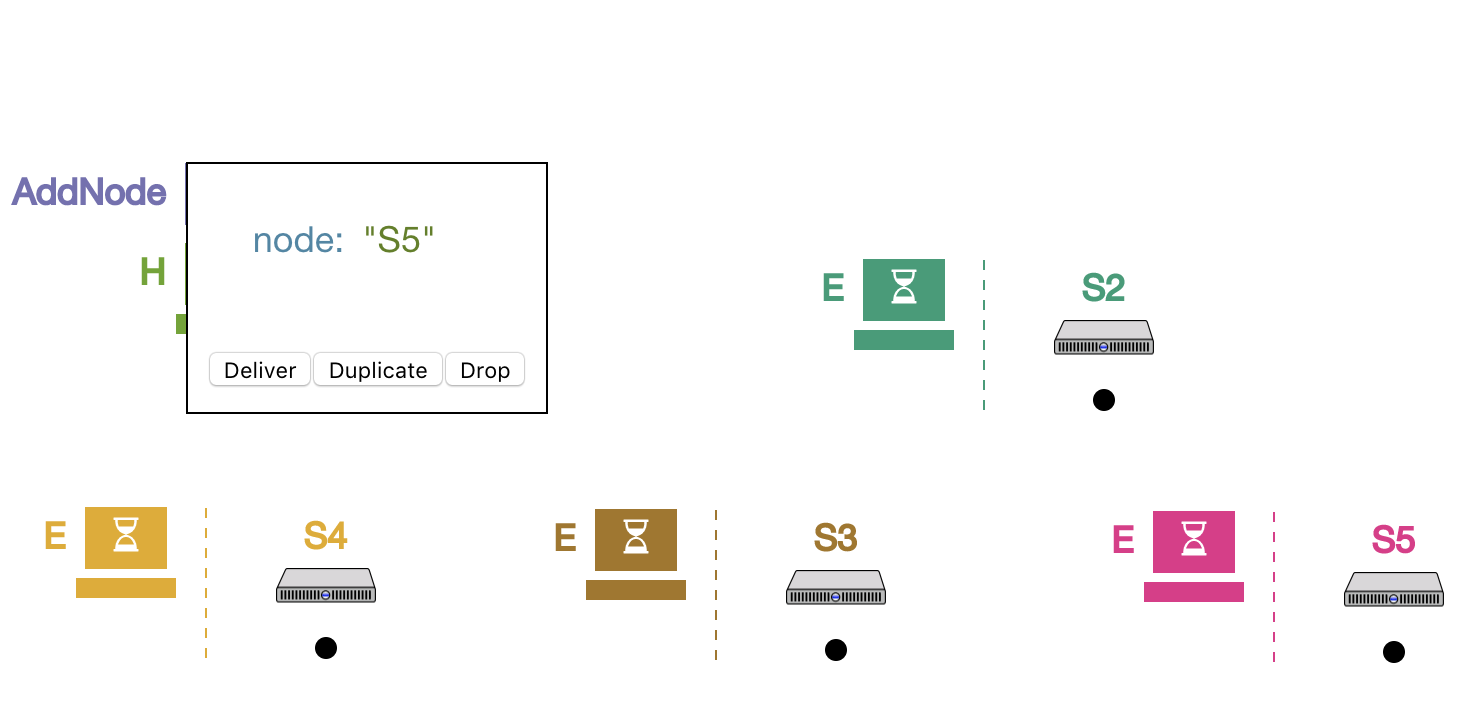}}
\caption{\label{fig:raft-4}The state of the Raft system in Oddity after client
  sends its reconfiguration request. The request is open for inspection. As
  shown, the engineer can choose to duplicate it or drop it rather than
  delivering it.}
\end{figure}

Now that $S_1$ is the leader, the engineer can investigate the reconfiguration
bug. They can make the client (not shown) send a reconfiguration request to add
$S_5$ by delivering a timeout. They can inspect the request by clicking on it,
as shown in Figure~\ref{fig:raft-4}. Once the request is delivered, the leader
will try to commit this new configuration to a majority of the new configuration
per the single-node reconfiguration protocol.

Once the new configuration has been replicated to $S_5$, the engineer needs to
trigger a new leader election in order to continue following the
counterexample. They can do so by delivering the \texttt{E} timeout to
$S_2$. This timeout is reset every time a heartbeat comes in from the leader,
and if it ever arrives the follower decides the leader has failed and makes
itself a candidate.

The rest of the leader election is elided for brevity. The engineer now clicks
on the client's other timeout, causing it to send a second reconfiguration
request dropping $S_1$. It is delivered at $S_2$ and $S_2$ replicates the new
configuration to $S_3$; this configuration is now committed, having been
replicated to a majority of the new cluster.

The counterexample now calls for $S_1$ to start a new election, which it can do
after receiving $S_2$'s \texttt{RV} message and then its \texttt{E}
timeout. After getting elected, it replicates the configuration with $S_5$ to
the rest of the cluster. Crucially, $S_1$ replicates the updated configuration
to $S_2$, overwriting a previously-committed entry and demonstrating that the
engineer's Raft implementation is buggy.

\subsection{Backtracking}

Investigating the Raft reconfiguration bug in Oddity involves a number of
steps. The engineer might mistakenly click on the wrong message (for instance,
delivering the node removal request to $S_2$ before the leader election
happens). The engineer might also want to explore alternative executions. The
reconfiguration bug can be fixed by requiring that a leader replicate an entry
(which could be a no-op) in its term before attempting to reconfigure the
system. In order to test this potential bug fix, the engineer could explore an
execution in which $S_2$ does this, attempting to replicate a no-op entry in its
old configuration before it receives the request to reconfigure the system. It
would be time-consuming and frustrating to start over from the initial state of
the system in order to answer such questions.

Fortunately, Oddity provides an alternative: the engineer can click on any
previous state in the history in order to reset the system to that state. They
can then explore other executions starting from that state. Using Oddity's
execution history view, the engineer can go back to the point just before $S_2$
started to replicate the command removing $S_1$ and instead deliver a heartbeat
timeout to $S_2$, causing it to attempt to replicate a no-op entry in the old
configuration.  In order to proceed, $S_2$ must replicate the no-op entry to at
least three nodes (e.g., $S_2$, $S_3$, and $S_4$) before it can attempt to
remove $S_1$ from the replica set. At that point the pending reconfiguration
with $S_5$ will not succeed, since $S_1$ will not be able to be elected leader
until its log is up to date with the rest of the cluster. The engineer now has
some evidence that the Raft reconfiguration bug can be fixed by requiring that a
leader replicates an entry in its term before attempting to reconfigure the
cluster.

%% file: frontend.tex
\section{Frontend}\label{sec:frontend}

\begin{figure}[!t]
\centering
\includegraphics[width=.9\linewidth]{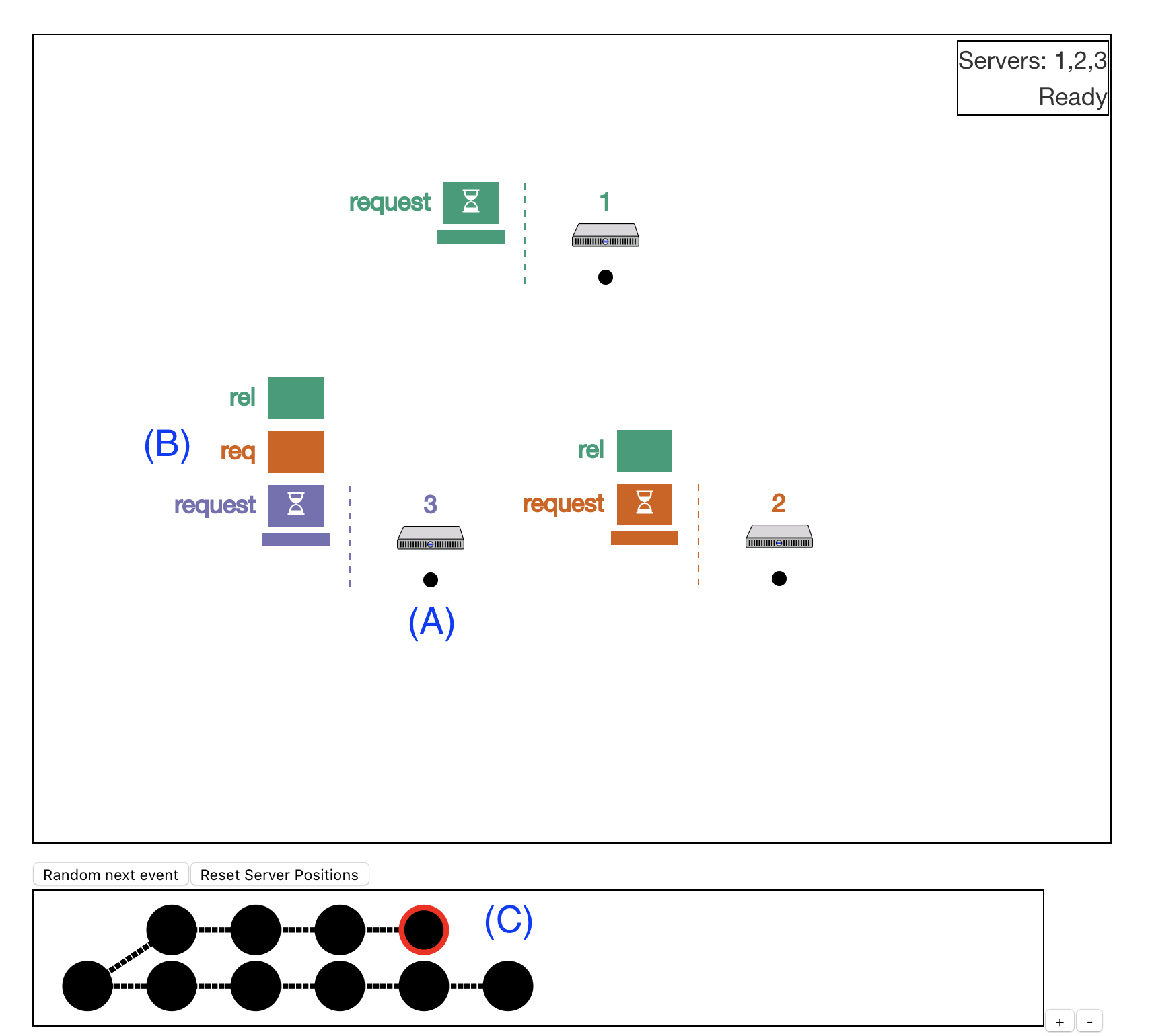}
\caption{\label{fig:debugger}The debugger window. Each node (A) is displayed,
  along with an inbox (B) of messages and timeouts waiting to be delivered at
  that node. The user can control delivery by clicking on timeouts and messages,
  and can also inspect the contents of any message or timeout or the state at
  any node. Using the branching history view (C), the user can navigate the
  states of the system they have explored. The user can reset the debugger to a
  previous state by clicking on it; this resets the system to that state so that
  the user can explore further from there.}
\end{figure}

Oddity's graphical interface, shown in Figure~\ref{fig:debugger}, is designed to
enable engineers to easily explore executions of distributed systems, including
failure cases. The graphical interface was designed with several requirements in
mind:
\begin{enumerate}
  \item It should be application- and implementation language-agnostic. A user
    should be able to graphically debug their system without developing a
    system-specific visualization.
  \item It should neither depend on nor suggest to the user any notion of real
    time. Messages can be arbitrarily delayed and reordered, and timeouts can
    be delivered even if no failures occur.
  \item It should enable detailed inspection of a single global state of the
    system (including the contents of all messages and the local state at every
    node), control over which event should be executed next, and navigation
    between system states for the purpose of time travel.
\end{enumerate}
In this section, we discuss how Oddity's frontend addresses each of these
requirements.

\subsection{Application-agnostic}

Distributed systems are designed to provide reliable service in diverse contexts
and this is reflected in their structure and operation: Chord~\cite{chord} and Dynamo~\cite{dynamo}
maintain a ring structure via pointers at each node, Raft has a single leader
who communicates with a number of followers, the DNS system has a loose tree
structure, etc. Rather than forcing system developers to develop visualizations
for the structure of each system, Oddity's interface displays the components all
distributed systems have in common: a set of nodes communicating via a
network. When the user starts Oddity, it displays each system node in a
circle. The user can then reposition the nodes as they desire by clicking and
dragging.

Oddity does not yet support application-specific extensions to the visualization
(for instance, to display a star next to Raft leaders or arrows describing
Chord's ring structure), but we anticipate that these will be easy to add. Thus
far, we have focused on making Oddity useful even for developers who are not
willing to develop such visualizations.

\subsection{No real time}

The unreliability of physical clocks due to clock skew is a fundamental problem
in distributed systems~\cite{Lamport:1978:TCO:359545.359563}. Engineers cannot
rely on measurements of time being consistent across nodes except within very
loose bounds. As a result, most distributed systems are designed around the
possibility that messages can be arbitrarily delayed by the network. While
messages are often thought of as moving through the network over time to their
destination, Oddity does not represent them this way; doing so would imply a
semantic meaning to real time. Instead, messages are immediately transferred to
the receiver's inbox and can then be delayed for an arbitrary amount of time (or
dropped), under user control. Oddity's display encourages users to ignore
wall-clock time in thinking about distributed systems correctness, and instead
think about correctness in the face of all possible event orders.

\subsection{State inspection and history navigation}

Oddity's graphical interface is geared towards representing a single state of
the system---including in-flight messages and timeouts---in detail, while also
enabling users to navigate a branching execution history. Users can click to
inspect server state or the contents of messages and timeouts. Enabling detailed
inspection is crucial for a debugging interface, since engineers use this
information to decide which message or timeout should be delivered next.  Oddity
supports time travel debugging, allowing engineers to navigate to any previously
explored state and explore a branching execution history without starting from
scratch.

%% file: backend.tex
\section{Architecture}\label{sec:backend}

\begin{figure}[t]
  \centering
  \includegraphics[width=.9\linewidth]{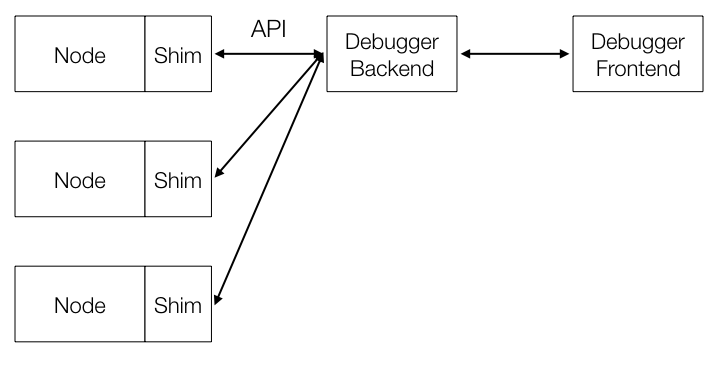}
  \caption{The architecture of the debugger implementation. The debugger backend
    communicates with individual nodes (each of which has a stub implementing
    the Oddity API). The debugger frontend, running in the browser, communicates
    with the backend. Most of the logic runs in the browser, allowing the
    backend to serve as a thin communication layer.}\label{fig:architecture}
\end{figure}

\begin{figure}[t]
  \centering
  \includegraphics[width=.9\linewidth]{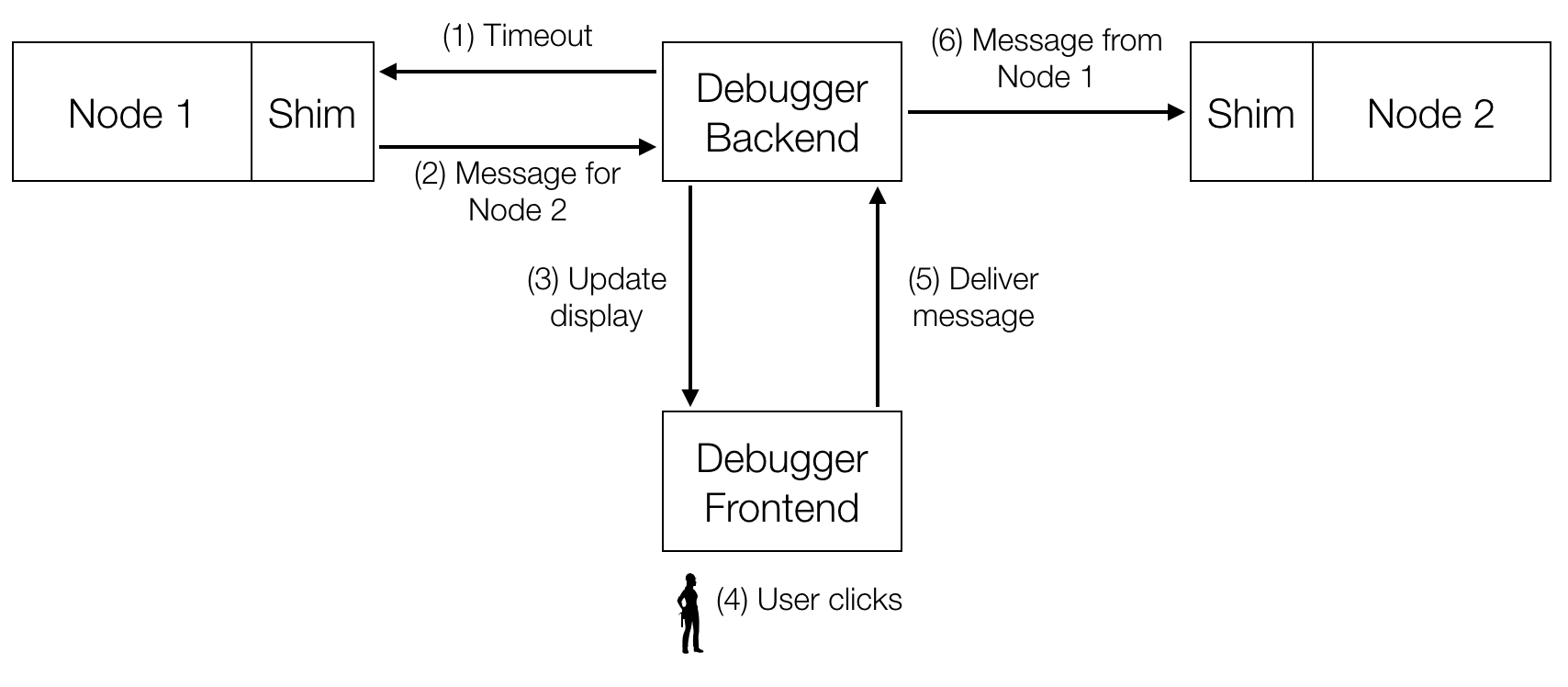}
  \caption{An example of DVIZ components communicating to deliver a
    message. When Node 1 receives a timeout, either from log replay or user
    action, it produces a message for Node 2. This message is sent to the
    debugger backend, which tells the frontend to display it in Node 2's
    inbox. When the user clicks the message (which they could do immediately or
    after delivering other messages and timeouts) the frontend notifies the
    backend, which sends the message to Node 2.}\label{fig:dataflow}
\end{figure}

\begin{table*}[h!t]
\caption{\label{tab:api}The Oddity API. In order to use Oddity, users must implement
  a simple, JSON-based message API. Once a system node registers with the
  server, it responds to each message (including the start message, which is
  sent at the beginning of a debugging session and after a reset) with its
  updated state, sent messages, and set and cleared timeouts.}  \centering
\begin{tabular}{l l l}
\textbf{Message} & \textbf{Description} & \textbf{From/To}\\
\hline
\verb~register(name)~ & Register a node & Node to server\\
\verb~start~ & Start the node & Server to node\\
\verb~timeout(type, body)~ & Deliver a timeout & Server to node\\
\verb~message(from, type, body)~ & Deliver a message & Server to node\\
\verb~response(state, messages, timeouts, cleared)~ & Response to any event & Server to node\\
\end{tabular}
\end{table*}

Figure~\ref{fig:architecture} shows the architecture of the Oddity debugger. As
shown, Oddity consists of several cooperating components: a browser-based
frontend that displays the interface discussed in the previous section; a
backend, split between the browser and the server, that tracks the system's
state, implements time travel, and communicates with nodes in the system; and a
shim that runs at each system node, communicates with the server, and calls the
node's handlers. Oddity's graphical interface is independent of its backend;
either could be replaced without changing the other.

\subsection{Debugger backend}

The debugger backend tracks the state of the system, including the local state
at each node and in-flight messages and timeouts. It also tracks the event
history. Each event is either a message delivery or a timeout delivery; a
special ``start'' event represents the beginning of time. When the user clicks
on a message or a timeout, the backend records this event and then sends the
message or timeout to the Oddity shim instance running on the appropriate
node. When it receives the response, it updates the display to reflect the new
messages and timeouts and the modified local state.

When the user clicks on a state in the history display, the debugger backend
resets the system to that point by replaying all of the events that led to that
state (including the special ``start'' event). The user can then explore
alternative executions starting from that state. This technique will not work if
the system has non-deterministic handlers or accesses persistent state outside
the debugger's control. Oddity could be extended to support such systems by
recording a snapshot of each node's state after every event. We leave such an
extension for future work.

\subsection{Oddity shim}

The Oddity shim is a library written in the user's implementation language that
implements communication with the Oddity server and is responsible for tracking
the local state of the system. It must implement the API shown in Table~\ref{tab:api}
for communication with the Oddity server, and must call the user's event
handlers for \texttt{start}, \texttt{timeout}, and \texttt{message} calls from
the server. It must provide user code with some mechanism for updating the local
state of the node, as well as sending messages and setting timeouts.

The Oddity shim currently assumes that the user's code is written as a set of
event handlers. Oddity cannot currently be used to debug systems developed
against a different programming model, such as multi-threaded servers using
blocking RPC calls. This limitation is not fundamental, and Oddity’s
architecture and visualizations can be extended to support such systems. We
discuss this potential future extension in more detail in
Section~\ref{sec:discussion}.

\subsection{Trace display}

Model checkers have been applied to distributed systems to great
effect~\cite{modist,macemc}. The counterexamples they produce, however, can be
long and complex. Rather than starting Oddity on a system and debugging from the
beginning of time, a user can start Oddity with a trace of that system's
execution, produced by a model-checker as a counterexample to a desired
property. A common use pattern is for the engineer to use time-travel to explore
the sequence of events that led to the invariant violation. The user can also
use Oddity to investigate other executions branching off of the trace. The same
approach facility can also be used to investigate traces mined from system logs
by tools such as DEMi~\cite{Scott:2016:MFE:2930611.2930631} to investigate bugs
encountered during testing or production use. Oddity enhances techniques such as
model-checking and log analysis by providing the engineer with tools to explore
the context of the bug as well as alternative executions that might or might not
trigger the same problem.

%% file: implementation.tex
\section{Implementation}\label{sec:implementation}

Our research prototype of the Oddity debugger is implemented in approximately
1400 lines of Clojurescript (for the browser-based frontend) and 500 lines of
Clojure (for the backend). Its interface uses SVG, and we have not found
rendering performance to be an issue even with large systems and long execution
traces. The current user interface is more limited---it is intended to be used
with roughly 10 or fewer nodes for regular debugging and exploration tasks. The
frontend uses a Websocket to communicate with the backend, which communicates
with the shim over TCP.

\begin{table*}[h!t]
\caption{\label{tab:systems} Systems that have been debugged using Oddity,
  the version of the shim they used, and lines of code for the system implementation.}  \centering
\begin{tabular}{l l l}
\textbf{System} & \textbf{Shim Language} & SLOC \\
\hline
Lamport mutual exclusion & Python & 73 \\
Raft (with reconfiguration) & Python & 240 \\
At-most-once RPC & Java & 280 \\
Primary-backup replication & Java & 380 \\
Paxos replication & Java & 550 \\
Sharded transactional key-value store & Java & 1390\\
\end{tabular}
\end{table*}

We have implemented the Oddity shim for Python (121 lines of code) and Java (293
lines of code), and we have developed several systems against each
implementation. It is easy to develop a new shim in any language with libraries
for JSON serialization and network communication. The course labs discussed in
the following section use the Java implementation of the
shim. Table~\ref{tab:systems} shows all of the systems that have been debugged
using Oddity, which version of the shim they used, and approximate lines of code
for each system. The largest system we have debugged using Oddity is a sharded
linearizable key-value store using Multi-Paxos replication to provide
exactly-once, highly available operations on each key. The shard allocations are
dynamic, and the system also supports simple multi-key transactions.

%% file: eval.tex
\section{Evaluation}\label{sec:eval}

We have deployed Oddity to two classes: a 40-student graduate-level
distributed systems class, which served as a pilot, and a 180-student
undergraduate-level distributed systems class. In both cases, Oddity was given
to students as part of the lab framework they used to do their homework
assignments. The labs come with extensive test suites, and include a
model-checker. Students can run Oddity in two modes: they can start their system
and explore from the beginning of time, or run Oddity on a counterexample trace
produced by the model-checker when an invariant is violated.

We had two goals in studying students' experience with Oddity. One was to
determine whether Oddity's features are useful. The other was to examine student
behavior when given access to an interactive debugger for distributed systems,
in line with previous work that examines student usage of traditional
step-through debuggers~\cite{doi:10.1080/08993400802114581} and developer usage
of trace visualization tools~\cite{trace-viz}. We hope that our experiences can
inform future work in the same area.

We studied student experiences with Oddity in several ways. We instrumented the
Oddity interface in order to track users' clicks on various interface elements
in order to see how they interacted with the system (this feature was only
enabled if students opted into it). We sent out an optional survey to students
after they completed the first major lab assignment, a primary-backup-based
key-value store (as of this writing, other labs are ongoing). The survey is
shown in Figure~\ref{fig:survey}. We also informally discussed Oddity with
students, recording anecdotes about their usage of the system on the
primary-backup lab as well as the next lab, a Paxos-based key-value
store~\cite{paxos}.

We have defined several research questions, each based on a different use-case
for Oddity: exploration of a system from the beginning of time in order to
understand its operation and to find possible bugs; diagnosing and understanding
a particular known bug; and replaying a trace from a model-checking
counterexample. The research questions are as follows:
\begin{description}
  \item{RQ1:} Do developers explore their systems starting from the beginning of time?
    When doing so, do developers use Oddity to test their systems' edge case behavior?
  \item{RQ2:} Do developers find the debugger useful in diagnosing and repairing bugs?
    When doing so, do developers explore multiple branches?
  \item{RQ3:} Is the ability to explore alternative executions starting from a
    model-checking trace useful for understanding why the bug occurred?
\end{description}
We discuss each of these questions in detail below. The two modes in which a
student can start the Oddity debugger are (1) to start it on their system and
explore from the system's start state and (2) to start it on a trace generated
by the model-checker when it finds a counterexample to a desired invariant. We
use these as rough proxies for (A) exploratory testing, in which developers
explore systems in order to understand them and find bugs and (B) diagnosing a
specific bug, respectively; this is imperfect, since students may start their
system from the beginning of time but with a specific bug in mind.

\para{RQ1} Do developers explore their systems starting from the beginning of
time?  When doing so, do developers use Oddity to test their systems' edge case
behavior?
We found that 74.5\% of Oddity runs started from the beginning of
time, as opposed to from a model-checking trace. In these runs, users explored
an average of 37.3 states per run, with a median of 23 states per run.  From this we can conclude that students did use the debugger for exploratory testing.
We received some survey data to suggest that students were able to explore edge
cases using this mode. One student said that \textit{\blockquote{It was useful
    for one bug where I found out there was unexpected behavior from the [view
      server] when both the primary and backup timed out at the same time.}}
This indicates that students used Oddity to explore edge case behavior.  We also
received an anecdote from a student about a bug in which their Paxos
implementation sent redundant messages under certain conditions (specifically:
when a proposer received more than a majority of replies to its ``prepare''
messages, it ended up sending extra ``accept'' messages). The student did not
suspect the existence of this bug before noticing it in Oddity, and believed
they would not have found the bug at all without Oddity (the provided test suite
did not test for the presence of these extra messages). Without an interactive
debugger that can control message and timeout ordering, exploratory testing of
distributed systems is tedious, and the usage tracking and survey
results indicate that students find this feature very useful.

A number of students said that they did not explore their systems starting from
the beginning because they only debugged their system when a test case from the
provided test suite failed. Our results may be biased as a result of our
setting: with an extensive test suite, some students may not have felt a need to
understand their system behavior independently of its behavior on the tests. It
is possible that without such an extensive test suite, students would have found
it more useful to start their systems in the debugger.
On the other hand, our evaluation is of students and not professional
developers. The students were learning about the protocols at the same time as
they were implementing and debugging them, so it is possible that exploratory
testing was a more compelling option for students than it would be for more
experienced developers.

\para{RQ2} Do developers find the debugger useful in diagnosing and repairing
bugs?  When doing so, do developers explore multiple branches?
We found that 25.5\% of Oddity runs started from a model-checking trace. In
response to survey question 2, students reported that Oddity \textit{\blockquote{helped
  [them] diagnose [their] handling of state transfer and state transfer
  acknowledgements}} and that they were able to use it to diagnose a bug in which
they \textit{\blockquote{had some delayed messages arriving and causing problems.}} A
student reported successfully using Oddity to diagnose a bug in which the system
had stopped making any progress after their latest change, which implemented
deduplication of redundant client requests. They stepped through a simple test
case and found that servers were never actually responding to clients; they were
then able to fix the issue.

\para{RQ3} Is the ability to explore alternative executions starting from a
model-checking trace useful for understanding why the bug occurred?
When students started their systems from a model-checking trace, 23.6\% of those
executions explored multiple branches. In those cases, those state graphs
branched an average of 1.5 times.  From this we can conclude that at least some students explored
alternative executions when viewing a model-checking counterexample. In response
to survey question 3, some students did report that exploring alternative
executions was useful. One student said that \textit{\blockquote{from the bug where our
  servers were advancing themselves based on outdated/future view numbers,
  instead of just from the view server, it helped us see a situation where we
  could get stuck more frequently waiting for the server to ack a state
  transfer.}} Another reported that the ability to explore alternative executions
\textit{\blockquote{distinctly helped understand what was happening.}} We can conclude
that the ability to explore alternative executions starting from a
model-checking counterexample was useful for some students.

\begin{figure}[t]
  \centering
  \begin{enumerate}
  \item Did the debugger help you to discover any bugs in your system? Describe one.
  \item Did the debugger help you to diagnose any bugs you were already aware of? Describe one.
  \item When using the debugger to view a search-test counterexample, did you also
   explore other executions? Did this help you to understand the
   counterexamples? Describe an instance of this being useful.
 \item Were there any bugs you think you would have found earlier if you had used
   the debugger? If not, how could the debugger have been more useful to you?
 \item Do you have any other feedback about the debugger?
  \end{enumerate}
  \caption{The optional survey sent to students after completing a homework
    assignment. We referred to tests that called the model checker as ``search
    tests.''}\label{fig:survey}
\end{figure}

%% file: discussion.tex
\section{Discussion}\label{sec:discussion}

Oddity provides an extensible platform for investigating many aspects of
distributed systems beyond the examples illustrated in earlier sections.
Below we describe how Oddity could facilitate new tools and
methodologies for debugging, implementing, and designing
distributed systems. We have left these features for future work.

\subsection{Interactive space-time diagrams}
\label{sec-6-1}

In addition to the primary ``nodes and inboxes'' visualization, Oddity
supports visualizing a system's execution as a space-time diagram (e.g.,
Figure~\ref{fig:raft-bug}).
Space-time diagrams are useful for viewing a summary of an entire execution
trace at once.
Because of Oddity's extensible design, adding a traditional (non-interactive)
version of space-time diagrams required less than 150 lines of code: Oddity
simply pipes a formatted version of the system trace through
GraphViz~\cite{graphviz}, and displays the resulting SVG image in the
browser.

In Oddity, space-time diagrams could be enriched with more detailed
information about the currently executing system, e.g., by adding JavaScript
hooks so that when a user hovers over a node in the space-time SVG, the state
of that individual node at that point in history is displayed.
Such an enriched space-time diagram would provide a bridge between the ``nodes and
inboxes'' and branching trace history visualizations.
Adding these additional features introduces new design and user interaction
challenges:
\begin{itemize}
  \item How should in-flight messages and timeouts be represented?
\item How should users interactively control system execution or time travel
from such a diagram?
\item In what scenarios is one visualization simpler or more effective than
another?
\end{itemize}
Oddity is well-suited for exploring these challenges: the Oddity API abstracts
away many of the tedious details for modeling the network, controlling
implementations of nodes, and interacting with different programming languages.

\subsection{System-specific interface components}
\label{sec-6-3}

The Oddity frontend is built around a generic SVG-based canvas which makes
integrating other visualization tools straightforward (e.g., for space-time
diagrams as discussed above).
In particular, Oddity could easily support systems which control aspects of
their own visual representation by providing a mechanism to add additional
shapes to the frontend SVG.
This could be as simple as nodes (optionally) providing a special field in
their local state with literal SVG objects to add to the visualization relative
to the node's own position.
These extensions would enable generic system-specific visualization.
For instance, the developer of a state-machine replication system might want to
display the log of commands seen at each node as an array of boxes colored by
term, while the developer of a ring maintenance system such as Chord might want
to display the successor and predecessor of each node as arrows to other nodes.
This would involve adding an API for systems to write elements to Oddity's
SVG-based interface, and perhaps developing a library of commonly-useful
components (such as the arrows mentioned above).
In general, Oddity's architecture makes integrating new visualizations easy.
Oddity's browser-based frontend also simplifies building on recent advances in data
visualization libraries such as D3~\cite{Bostock:2011:DDD:2068462.2068631}.

\subsection{System model}
\label{sec-6-2}

Initially, we have focused on using Oddity to debug and explore distributed
systems where node implementations behave as deterministic, single-threaded
event handling loops.
This class of systems corresponds to the example protocols used in lectures and
exercises for the courses where Oddity has been used to date
(Section~\ref{sec:eval}). It also corresponds to the model used by recent projects formally verifying implementations
of distributed systems~\cite{verdi, ironfleet, chapar}.
However, many distributed systems rely on some combination of nondeterministic
choice, blocking RPC-based communication, and local multi-threading at nodes.

For basic control and visualization, Oddity already supports nondeterministic
distributed systems.
However, Oddity's replay-based approach to time travel will not correctly
restore local node states with nondeterministic handlers.
To correctly provide time-travel in the face of nondeterminism, Oddity could
require that each node send a snapshot of its current state after each event,
or we could extend the Oddity shim to fork a child process after each event to
effectively save a ``paused'' instance of a node's process.
Equipped with such an extension to the backend (and without any
changes to the frontend), Oddity could support time travel for nondeterministic
systems by restoring arbitrary previous states from snapshots.

Many implementations of distributed systems allow for the concurrent execution
of event handlers for higher performance; for example, updates to different
keys in a key-value store can be safely handled in parallel on different 
processors of a multi-core server, e.g., using locks to arbitrate access
to shared data structures.  Thus, the system behavior may depend on
the thread scheduling decisions made on the local node.  If the bug being
diagnosed is invariant to the local thread schedule, it may suffice to simply
enforce a single canonical thread execution order, such as with deterministic
multi-threading~\cite{dmp}.  If the bug manifests due to the interaction of the local
schedule and distributed event delivery, then we would need to extend 
the visualization model to allow the programmer to control both. 

Supporting RPC-based systems adds another layer of complexity to 
the interface between the debugger and the system.
Oddity's system model assumes that nodes atomically send messages in response
to receiving a message or timeout and then immediately return to the top of the
event loop, ready to handle the next (arbitrary) input event.
With RPC systems, the event loop is inverted. Although the code performs
the same computation in the same order as in an event system, 
a message arrival is ``handled'' only when a thread retrieves it, e.g., in
response to a previous send.

\subsection{Deeper model-checker integration}
\label{sec-6-4}

Oddity is designed to use event traces as a common representation for
communication between the frontend and backend and between the backend and node
shims.
This design choice made integration with a model checker straightforward: the
backend can simply walk a counterexample produced from the model checker as a
trace and use each step to replay events on the nodes as in time travel
debugging.
As discussed in Section~\ref{sec:eval}, this straightforward technique for
integrating model checkers has already proved valuable for Oddity users.
To further integrate model-checker functionality, Oddity could highlight
particular components of a system state that violate the desired invariant.

More ambitiously, a model-checker could also be used to provide Oddity with
``breakpoints.'' Many systems do some initial bootstrapping and setup that may
be tedious to simulate manually in Oddity (for instance, electing an initial
leader). Instead, a developer could specify that they want to debug the system
starting in some state meeting a global property (such as after a successful
election).
Oddity could ask the model-checker to find such a state, and then allow the
user to explore the system's execution starting from the state returned by the
model-checker.

%% file: related.tex
\section{Related Work}\label{sec:related}
\label{sec-3}
Oddity builds on previous work in several areas, including distributed systems correctness,
distributed systems log exploration, and distributed systems visualization. We
discuss each of these areas in detail below. In addition, Oddity depends on a
long line of work on single-node program debugging; many aspects of Oddity's
design were inspired by graphical step-through debuggers such as DDD~\cite{ddd}.

\paragraph{Distributed Systems Correctness}
Systems such as TLA+~\cite{tla}, Alloy~\cite{alloy}, and Ivy~\cite{ivy} have
been used for bug-finding and verification of high-level, abstract models of
distributed systems (Ivy has recently been extended to support extraction to
runnable code~\cite{ivy-pldi}). All
three share Oddity's goal of enabling users to understand the behavior of their
systems and encouraging correct distributed systems thinking. TLA+ and Alloy use
bounded model-checking; Oddity could be used in conjunction with these systems
to display model-checking counterexamples.  Ivy enables automatic verification
of distributed systems specified in a carefully-crafted subset of first-order
logic; it includes a graphical representation of counterexamples. Oddity
complements these systems by enabling users to interactively debug
both models and working implementations of distributed systems during
development and after deployment.
\label{sec-3-1}
\label{sec-3-2}
\paragraph{Distributed systems log analysis}
There has been a large amount of work on collecting and analyzing logs of
distributed
systems~\cite{Bates:1983:AHD:800007.808017,Eick:1996:IVM:525394.837855,Kranzlmuller:1996:EGV:238020.238054,380478,Kunz97poet:target-system-independent,Zernick:1992:UVT:624593.625178,d3,shiviz}
and, relatedly, datacenter
networks~\cite{Mai:2011:DDP:2043164.2018470,Tammana:2016:SDN:3026877.3026896}. Many
of these systems, such as ShiViz~\cite{shiviz}, provide a graphical interface,
allowing users to interactively explore the logs produced by their
system. ShiViz's visualization is based on space-time diagrams; it allows users
to explore large and complex executions by querying the log and collapsing and
expanding events. Like Oddity, these tools are designed for debugging and
understanding distributed systems implementations. Unlike Oddity, these tools
are for ex post facto debugging of system logs, rather than interactive
debugging of a system's behavior; log analysis systems do not enable exploratory
testing. These two use cases complement each other: having obtained and examined
a system log using ShiViz, an engineer can replay the log locally using Oddity
in order to understand it and diagnose any problems that were encountered. Using
Oddity, a user can explore alternative message orderings to determine whether
they also produce bugs; existing log analysis systems do not support this.

\paragraph{Distributed systems animations}
Runway~\cite{runway} is a system for visualizing models of distributed
systems. It consists of a domain-specific high-level modeling language based on
TLA-like actions. An interpreter for this language is written in
Javascript and an API for extracting values from the interpreter for
visualization. Several models and animations have been developed using Runway,
including a visualization of the Raft consensus protocol.

Oddity's visualization was inspired by those created for Runway.  Oddity is much
more general, however. Users of Runway must create protocol-specific
visualizations, e.g., there is a visualization for Raft that would not apply
even to related protocols such as Multi-Paxos or primary-backup. The students to
whom we gave Oddity would have had to write their own Runway visualizations,
since they were instructed to implement against a specification but were not
forced to use any particular algorithm. Additionally, Oddity can be used to
debug systems written in any language, while Runway requires users to model
their systems in its domain-specific language.

%% file: conclusion.tex
\section{Conclusion}\label{sec:conclusion}
\label{sec-6}

Oddity is the first interactive graphical debugger for distributed systems.
It allows users to control the delivery of events to the distributed system and
observe the resulting execution.
Oddity provides time travel to enable debugging multiple executions in order
to explore normal- and edge-case behavior.
Oddity introduces a new visualization and interaction mode that encourages
distributed systems thinking: rather than assuming the normal case and
attaching failure handling as an afterthought, users are shown the vast range
of possible behaviors and provided with the tools needed to effectively
explore.
Dozens of student users learning about distributed systems across graduate
and undergraduate courses have reported the value of such exploration when
debugging and learning what makes systems (not) work.
Finally, Oddity provides an extensible platform that can support research
investigating new questions about how best to visualize and explore a greater
range of systems.